\documentclass[twocolumn,english,pra,aps,superscriptaddress]{revtex4-2}
\usepackage[T1]{fontenc}%\usepackage[utf8]{inputenc}
\usepackage{color}
\usepackage{babel}
\usepackage{mathtools}
\usepackage{bm}
\usepackage{graphicx}
\usepackage{esint}
\usepackage{amsfonts}
\usepackage{float}
\usepackage{subfigure}
\usepackage{amsthm}
\usepackage{amssymb}
\usepackage{braket}
\usepackage{physics}
\usepackage[pdftex,colorlinks=true, linkcolor = blue, citecolor=blue,urlcolor=blue, 
bookmarksnumbered=true, bookmarksopen=true]{hyperref}
\usepackage{breakurl}
\usepackage{orcidlink}
\usepackage{listings}

\begin{document}

\title{Optimal noisy quantum phase estimation with finite-dimensional states}

\author{Jin-Feng Qin\orcidlink{0000-0001-8794-9793}}
\affiliation{Center for Theoretical Physics and School of Physics and Optoelectronic Engineering, 
Hainan University, Haikou 570228, China}
\affiliation{Mingde College, Chengdu Technological University, Yibin 644000, China}
\affiliation{School of Physics, Huazhong University of Science and Technology, Wuhan 430074, China}

\author{Jing Liu\orcidlink{0000-0001-9944-4493}}
\email{jing.liu@hainanu.edu.cn}
\affiliation{Center for Theoretical Physics and School of Physics and Optoelectronic Engineering, 
Hainan University, Haikou 570228, China}

\begin{abstract}
Phase estimation in quantum interferometry is a major scenario where the quantum advantage is significantly 
revealed. Recently, the optimal finite-dimensional probe states (OFPSs) for phase estimation in two-mode 
quantum interferometry have been provided with the absence of noise [J.-F. Qin et al., 
\href{https://doi.org/10.1103/953x-1rkm}{Phys. Rev. A \textbf{112}, 052428 (2025)]}. 
However, the noise is inevitable in practice and the previously obtained OFPSs may cease to be optimal anymore. 
Hence, the forms of the true OFPSs in the existence of various noises are still open questions.
Hereby, the noise of particle loss is studied and the true OFPSs under this noise have been investigated with the 
numerical algorithm named constrained optimization by linear approximation. Furthermore, a two-step measurement 
strategy is proposed to realize the ultimate precision limit in practice. The validity of this strategy is confirmed by 
the numerical simulation of practical experiments. 
\end{abstract}

\maketitle

\section{Introduction}

Phase estimation in quantum interferometry is the first scenario that proves the standard quantum limit 
can be overcome~\cite{Caves1980,Caves1981,Xiao1987}, and with years of development, it has become one of the 
most important topics in quantum metrology. To provide an optimal scheme for phase estimation, searching 
for the optimal probe state is always the first thing to do, and has certainly been given extensive attention  
in many years~\cite{Yurke1986,Bollinger1996,Giovannetti2006,Pezze2009,Genoni2011,Demkowicz2011,Humphreys2013,
Jarzyna2013,Lang2013,Sahota2015,Altenburg2016,Ragy2016,Lang2014}. The Fock basis, of which the elements are 
the eigenstates of the number operator, is a common representation of quantum states in many quantum systems, 
especially the bosonic systems. Many finite-dimensional states in the Fock basis are proved to be very powerful in 
quantum metrology~\cite{Sanders1989,Boto2000,Holland1993,Vogel1993,Park2017,Boas2019,Lang2014,Datta2011,
Luo2017,Boto2000,Mitchell2004,Nagata2007}, such as the well-known twin-Fock state~\cite{Holland1993,Lang2014,
Datta2011,Luo2017} and N00N state~\cite{Sanders1989,Boto2000,Mitchell2004,Nagata2007}. However, the consistency 
between the average particle number and state dimension in these states makes it difficult to distinguish the contributions 
of the average particle number and state dimension to the final precision. Hence, investigations on those finite-dimensional 
states with inconsistent average particle number and state dimension become an emerging topic~\cite{Shapiro1989,
Shapiro1991,Braunstein1994,Kowalewska2006,Genoni2007,Lee2016,Lee2019,Qin2025,Lu2022} in recent years, especially 
searching for the optimal ones in various scenarios~\cite{Lee2016,Lee2019,Qin2025,Lu2022}. 

Recently, Lee \emph{et al.}~\cite{Lee2019} provided the optimal finite-dimensional probe state (OFPS) with respect 
to a fixed average particle number for the phase estimation in a single-mode bosonic system. Thereafter, the 
OFPSs for estimating the phase difference in both linear and nonlinear two-mode quantum interferometries have been 
given and thoroughly discussed~\cite{Qin2025} with the absence of noise. Moreover, it has been shown that both 
parity~\cite{Gerry2007,Gerry2010a,Gerry2010b,Plick2010,Cohen2014} and particle-counting~\cite{Banaszek2003,Fitch2003,
Silberhorn2007,Zhong2018,Tiedau2021} measurements are optimal measurements for the OFPS at the point of true value. 
The dependence of the optimality of these two measurements on the true value can be well eliminated via the adaptive 
measurements, where a tunable phase is involved and tuned based on the sharpness or other quantities~\cite{Berry2000,
Berry2001,Bargatin2005,Dobrzanski2017,Garcia2022,Garcia2022,Kurdzialek2023,Albarelli2022,Sieniawski2026,Liu2022}. 
The proposed scheme with the OFPS provides a new aspect to enhance the measurement precision of phase difference 
without increasing the particle number of the probe. This scheme would be very useful in scenarios like biological 
detection~\cite{Taylor2013}, where a  weak light probe is required to avoid damaging the biological specimen, and in 
cost-effective environments including the satellites~\cite{Lu2022a} and chips~\cite{Stokowski2023}.  It would also be 
very useful in the scenarios where the noise floor is strongly related to the particle number. 

For the sake of providing a rigorous analytical result, the OFPS in the two-mode interferometry is first studied in the ideal 
scenario, namely, no noise is involved. However, the existence of noise is inevitable in practice. In quantum interferometry, the 
particle loss is one of the major noise modes and has been widely investigated in recent years~\cite{Huver2008,Demkowicz2009,
Lee2009,Dorner2009,Jiang2012,Liu2013,Knott2014b,Knott2016,Lee2020}. In the case that the particle loss exists, the found forms 
of the OFPS may cease to be optimal anymore, at least in mathematics. Therefore, what the true OFPSs are under particle loss and 
how they perform in practice are still open questions. Providing answers to these questions is the major motivation of this work. 

When the noise of particle loss exists, the analytical optimization of the probe state becomes extremely hard, if not fully impossible, 
since the state is mixed, and numerical optimization methodology has to be involved. In this paper, the constrained optimization by 
linear approximation (COBYLA) algorithm~\cite{Powell1994,Powell1998,Powell2007} is used as the numerical method to locate the 
noisy OFPS for different amounts of particle loss. In contrast with the noiseless scenario, the parity and particle-counting 
measurements are no longer optimal under particle loss, and in this case, the optimal measurement can be constructed via the 
eigenvectors of the symmetric logarithmic derivative. Unfortunately, this measurement is unrealizable in practice due to its dependence 
on the true value. To solve this problem, a two-step measurement strategy is proposed  and its validity in practice is confirmed by 
numerical simulations of the experiments. 
 
\section{Theoretical model and depiction of the precision limit}

A common way to accumulate phase in a single-mode bosonic system is usually realized by the operator 
$\exp(i\phi_a a^{\dagger}a)$ with $a$ the annihilation operator of the mode and $\phi_a$ the accumulated phase. 
In a two-mode bosonic system with $a$, $b$ the annihilation operators, the total phase accumulation for both 
modes is $\exp(i\phi_a a^{\dag}a+i\phi_b b^{\dag}b)$ with $\phi_b$ the accumulated phase on mode $b$. Utilizing 
the Schwinger operator $J_z=(a^{\dag}a-b^{\dag}b)/2$ and number operator $n=a^{\dag}a+b^{\dag}b$, the total 
phase accumulation can be rewritten into 
\begin{equation}
e^{i\phi_{\mathrm{tot}}n/2}e^{i\phi J_z}=:U_{\mathrm{lin}},
\label{eq:Ulin}
\end{equation}
where $\phi_{\mathrm{tot}}=\phi_a+\phi_b$ is the total phase and $\phi=\phi_a-\phi_b$ is the phase difference.
This type of phase shift is usually called the linear phase shift since the accumulation operator $\phi_a a^{\dagger}a$ 
is proportional to $a^{\dagger}a$. When the accumulation operator is proportional to the $k$th power of $a^{\dag}a$, 
i.e., $\phi_a (a^{\dag}a)^k$, the corresponding phase shift is usually called the nonlinear phase shift. In this paper 
we only consider the nonlinear case with $k=2$, which can be readily realized in the current experiments. The nonlinear 
cases with higher powers will be further investigated in the near future. When both phase shifts on two modes are nonlinear 
with $k=2$, the total phase accumulation is expressed by 
\begin{equation}
e^{i\phi_{\mathrm{tot}}[(a^{\dagger}a)^2+(b^{\dagger}b)^2]/2}e^{i\phi n J_z}=:U_{\mathrm{non}}.
\end{equation}
Although two phases $\phi_{\mathrm{tot}}$ and $\phi$ exist in the operators $U_{\mathrm{lin}}$ and $U_{\mathrm{non}}$, their 
absolute values cannot be directly measured simultaneously without an external reference. In the meantime, if the absolute 
value of $\phi_a$ or $\phi_b$ is needed to be measured, the absolute value of the other one has to be known in principle, which 
means the final precision of the unknown phase would be limited by the precision of the known phase. In this sense, the quantum 
advantage of the interferometry could be fully useless as long as the known phase is measured in a traditional way and its 
precision is limited by the standard quantum limit. Therefore, in this work we focus on the estimation of the phase difference 
$\phi$ which would avoid the aforementioned problems. Both linear and nonlinear phase shifts will be studied. 

%======================================= Figure ======================================
\begin{figure}[tp]
\centering
\includegraphics[width=8.5cm]{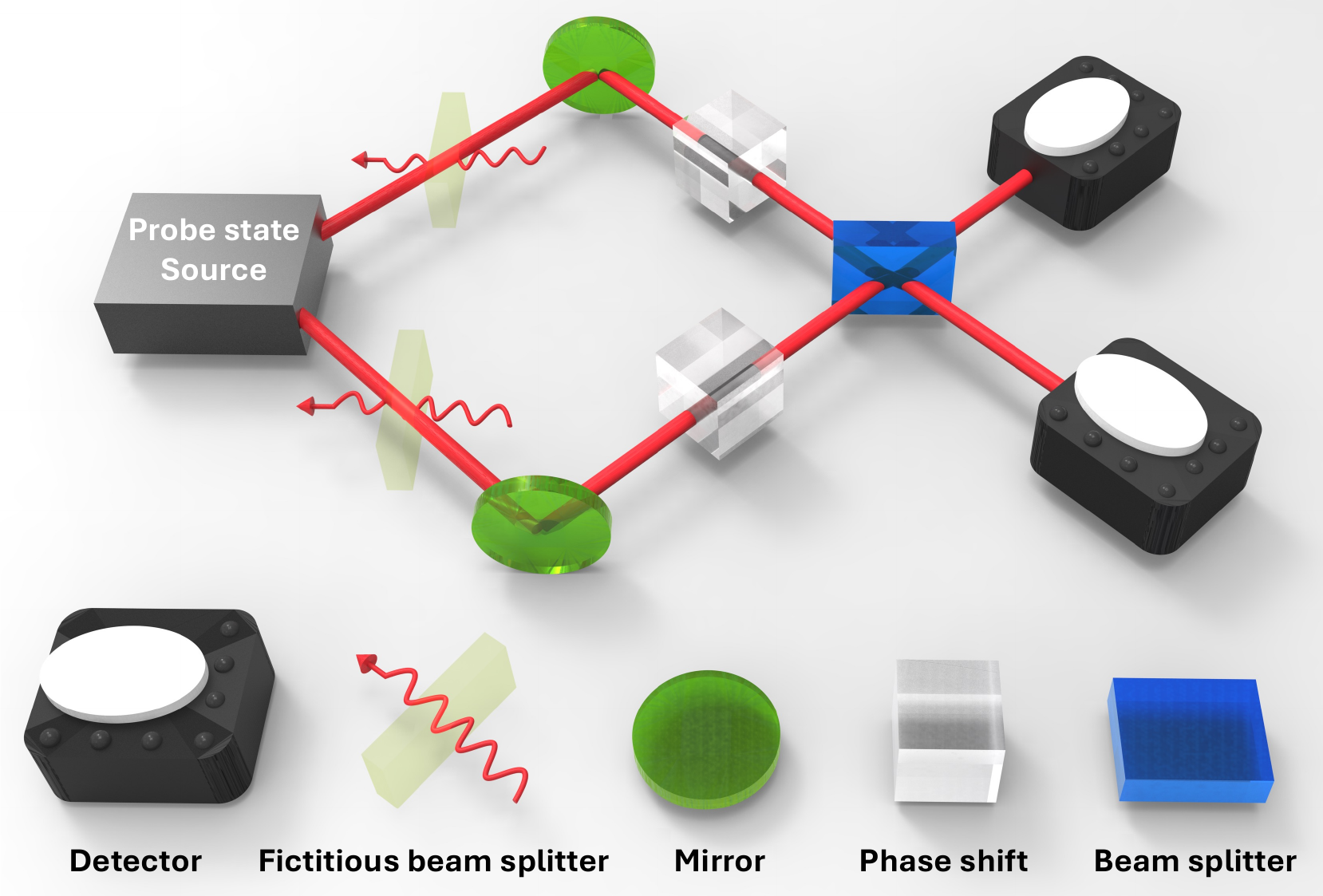}
\caption{Illustration of the phase estimation and particle loss in the scenario of 
the quantum optical interferometry. The light-green blocks represent the fictitious 
beam splitters, which are used to depict the particle loss in theory.}
\label{fig:schematic}
\end{figure}
%===================================================================================

Particle loss is a common noise mode in quantum phase estimation, which 
in theory can be modeled by the fictitious beam splitter~\cite{Huver2008,Demkowicz2009,Dorner2009}. 
In a two-mode bosonic quantum interferometric model, the particle 
losses on two modes are usually modeled by two fictitious beam splitters. Recall that the 
modes $a$ and $b$ are those of both arms. The operators for the fictitious beam 
splitters can be written as 
\begin{eqnarray}
B^{T_{1}}_{a c} &=& e^{i \frac{\eta_{1}}{2}(a^{\dag}c+ac^{\dag})},
\label{eq:fictitious_BS1}\\
B^{T_{2}}_{b d} &=& e^{i \frac{\eta_{2}}{2}(b^{\dag}d+bd^{\dag})},
\label{eq:fictitious_BS2}
\end{eqnarray}
where $c$ and $d$ are two fictitious modes representing the particle loss. $T_{1}=\cos^{2}(\eta_{1}/2)$ and $T_{2}=\cos^{2}(\eta_{2}/2)$
are the transmission coefficients. In the case that $T_1$ ($T_2$) equals $1$, no loss happens 
from the mode $a$ ($b$), and when $T_1$ 
($T_2$) equals zero, all particles are lost from the mode $a$ ($b$). It has been shown that~\cite{Huver2008,Demkowicz2009,Dorner2009} 
the results would be the same when the particle loss occurs before or after the state goes through the phase shifts. Hence, here we take 
the case that the particle loss occurs first, and in this case the state going through the phase shifts is actually a mixed state. To better 
present the physical scenario we discuss, the phase estimation and particle loss are illustrated in Fig.~\ref{fig:schematic} via the example 
of quantum optical interferometry. Notice that the light-green and blue blocks represent the fictitious and physical 
interferometer beam splitters, respectively. However, the main results of this work are not limited to this scenario. 

The full quantum state of all modes $a,b,c,d$ before the particle loss happens can be written as 
$\ket{\psi_{\mathrm{in}}}\otimes\ket{0}_{c}\otimes\ket{0}_{d}$, where $\ket{\psi_{\mathrm{in}}}$ 
is the probe state on modes $a$ and $b$. $\ket{0}_c$ ($\ket{0}_d$) is the vacuum state of the 
fictitious mode $c$ ($d$), and $\otimes$ represents the tensor product. 
Based on the fictitious beam splitter operators given in Eqs.~(\ref{eq:fictitious_BS1})
and (\ref{eq:fictitious_BS2}), the full state going through the fictitious beam splitters is 
\begin{equation}
\ket{\psi} = B^{T_{2}}_{b d}B^{T_{1}}_{a c} \ket{\psi_{\mathrm{in}}}\otimes\ket{0}_{c}\otimes\ket{0}_{d}.
\end{equation}
By tracing out the unobserved environmental modes $c$ and $d$, 
the reduced density matrix on modes $a$ and $b$ is 
\begin{equation}
\rho = \mathrm{Tr}_{c d}\left(B^{T_{2}}_{b d}B^{T_{1}}_{a c}\ket{\psi}\bra{\psi}
{{B^{T_{1}}_{a c}}}^{\dagger}{{B^{T_{2}}_{b d}}}^{\dagger}\right), 
\label{eq:rho}
\end{equation}
where $\mathrm{Tr}_{cd}(\cdot)$ is the partial trace on modes $c$ and $d$. The parameterized state then reads 
$\rho_{\phi}=U_{\mathrm{lin}}\rho U_{\mathrm{lin}}^{\dag}$ and $\rho_{\phi}=U_{\mathrm{non}}\rho U_{\mathrm{non}}^{\dag}$ 
for the linear and nonlinear phase shifts, respectively. 

In the quantum parameter estimation, the quantum Cram\'{e}r-Rao bound (also known as Helstrom bound) is one of the most 
well-used tools to depict the precision limit. In this theory, the precision of the phase difference is depicted by the variance 
$\delta^2\phi$. Furthermore, the variance satisfies the following inequality~\cite{Helstrom1976,Holevo1982}
\begin{equation}
\delta^2\phi \geq \frac{1}{\mu I}\geq \frac{1}{\mu F}.    
\end{equation}
Here $\mu$ is the repetition number of the experiment, and $I$ and $F$ are the classical Fisher information (CFI) and quantum Fisher 
information (QFI). For a set of discrete probability distribution $\{p_i\}$, the CFI can be calculated via the equation 
$I= \sum_i(\partial_{\phi}p_i)^2/p_i$. Here $\partial_{\phi}$ is short for $\partial/\partial \phi$. For the density matrix $\rho_{\phi}$, 
the QFI can be calculated via the expression $F=\mathrm{Tr}(\rho_{\phi}L^2)$~\cite{Helstrom1976,Holevo1982}. Here $L$ is the 
symmetric logarithmic derivative (SLD) and is determined by the equation $\partial_{\phi}\rho_{\phi}=(\rho_{\phi}L+L\rho_{\phi})/2$. 
 
Now denote the spectral decomposition of $\rho$ in Eq.~(\ref{eq:rho}) as $\rho=\sum_{\lambda_i\in\mathcal{S}}\lambda_i\ketbra{\lambda_i}$ 
with $\lambda_i$ and $\ket{\lambda_i}$ the $i$th eigenvalue and eigenstate of $\rho$ and 
$\mathcal{S}=\{\lambda_i\in\{\lambda_i\}|\lambda_i\neq 0\}$. According to the expression of the QFI for unitary parameterization 
processes~\cite{Liu2015,Liu2020,Liu2014}, the QFI for the linear phase shifts can be calculated by 
\begin{equation}
F=\sum_{\lambda_i\in S}\!\!4\lambda_i\mathrm{var}_{\ket{\lambda_i}}(J_z)-\!\sum_{\underset{i\neq j}{\lambda_i,\lambda_j \in S}}
\!\frac{8\lambda_i \lambda_j}{\lambda_i+\lambda_j}|\!\expval{\lambda_i|J_z|\lambda_j}\!|^2,
\label{eq:QFI}
\end{equation}
where $\mathrm{var}_{\ket{\lambda_i}}(J_z):=\expval{\lambda_i|J^2_z|\lambda_i}-\expval{\lambda_i|J_z|\lambda_i}^2$. For the nonlinear 
phase shifts, the expression of the QFI can be obtained by replacing $J_z$ with $n J_z$ in the equation above. In this work, the QFI will 
be used as the objective function for the optimization. 

\section{The noiseless OFPS}

The finite-dimensional states are well-studied in the quantum parameter estimation and constantly compared to the continuous-variable 
states, especially in the two-mode systems.  In the past, the study of two-mode finite-dimensional states usually focus on the form 
$\sum_{i=0}^{N} c_{i}\ket{i, N-i}$ with $\ket{i, N-i}$ a Fock state and $c_{i}$ a complex coefficient. The advantage of this form 
is that the average particle number is always fixed and can be readily realized in quantum systems like cold atoms in a double-well 
potential. However, the disadvantage is that the average particle number always equals to the dimension of the state minus one, 
and their contributions to the enhancement of the final precision cannot be distinguished. For the sake of investigating the effect 
of the state dimension on the precision, a more general form should be used, which reads  
\begin{equation}
\ket{\psi}=\sum_{i,j=0}^{N} c_{ij}\ket{i j}.
\label{eq:general_form}
\end{equation}
Here $N$ is referred to as the Fock dimension and $\ket{ij}$ is a Fock state. $c_{ij}$ is a complex coefficient. The average particle 
number (denoted by $\bar{n}$) for this state is 
\begin{equation}
\bar{n}=\sum^N_{i,j=0}|c_{ij}|^2(i+j),
\end{equation}	
which could be chosen as any value in the region $[0,2N]$. Hence, the contributions of the average particle number and the Fock 
dimension can be identified independently. 

Based on the form in Eq.~(\ref{eq:general_form}), the OFPS can be analytically or numerically obtained by maximizing the QFI. 
Recently, the OFPSs for this case have been analytically provided in the absence of noise~\cite{Qin2025}. The form of the OFPS 
relies on the region of $\bar{n}$. For the linear phase shifts, in the region $\bar{n}\in(0,N]$ the OFPS reads  
\begin{equation}
\sqrt{\frac{N-\bar{n}}{N}}\ket{0 0}+\sqrt{\frac{\bar{n}}{2N}}
\left(e^{i\theta_1}\ket{0 N} +e^{i\theta_2}\ket{N 0}\right),
\label{eq:lin_Optstate}
\end{equation}
where $\theta_1,\theta_2\in[0,2\pi)$ are the relative phases. In the region $\bar{n}\in[N,2N)$, the OFPS reads 
\begin{equation}
\sqrt{\frac{2N-\bar{n}}{2N}}\left(e^{i\theta_1}\ket{0 N}\!+\!e^{i\theta_2}
\ket{N 0}\right)\!+\!\sqrt{\frac{\bar{n}-N}{N}}\ket{N N}. 
\label{eq:lin_Optstate1}
\end{equation}
The QFIs for the states in Eqs.~(\ref{eq:lin_Optstate}) and (\ref{eq:lin_Optstate1}) are $\bar{n}N$ and $N(2N-\bar{n})$, respectively. 

For the nonlinear phase shifts, in the region $\bar{n}\in(0,N]$, the OFPS is still the state given in Eq.~(\ref{eq:lin_Optstate}).
Furthermore, in the region $\bar{n}\in\left[N, \left\lfloor\frac{4N+1}{3}\right\rfloor\right]$, the OFPS is 
\begin{align}
& \sqrt{\frac{\bar{n}-\lfloor \bar{n}\rfloor}{2}}
\left(\ket{\lfloor \bar{n}\rfloor\!+\!1\!-\!N, N}
\!+\!e^{i\theta_1}\ket{N, \lfloor\bar{n}\rfloor\!+\!1\!-\!N}\right)\notag\\
&+\sqrt{\frac{1-(\bar{n}-\lfloor\bar{n}\rfloor)}{2}}
\left(e^{i\theta_2} \ket{\lfloor \bar{n}\rfloor\!-\!N, N}
\!+\!e^{i\theta_3}\ket{N, \lfloor\bar{n}\rfloor\!-\!N}\right), 
\end{align}
where $\theta_1,\theta_2$, and $\theta_3$ are the relative phases. The symbol $\left\lfloor\cdot\right\rfloor$ represents the 
floor function. In this case, if $\bar{n}$ is a integer, the state above reduces to 
\begin{equation}
\frac{1}{\sqrt{2}}\left(\ket{\bar{n}-N, N}+e^{i\theta}\ket{N,\bar{n}-N}\right)   
\end{equation}
with $\theta$ a relative phase. Furthermore, in the region 
$\bar{n}\!\in\!\left[\left\lfloor\frac{4N+1}{3}\right\rfloor,2N\right)$, the OFPS is 
\begin{equation}
\sqrt{\frac{2N-\bar{n}}{2\left(N-\zeta\right)}}\Big(e^{i\theta_1}\ket{\zeta N}
+e^{i\theta_2} \ket{N \zeta}\Big)+\sqrt{\frac{\bar{n}-N-\zeta}{N-\zeta}}\ket{NN}.
\end{equation}
Here $\zeta = \left\lfloor\frac{N+1}{3}\right\rfloor$, and $\theta_1,\theta_2$ are the relative phases. 
If $N/3$ is an integer, the OFPS in this case reduces to 
\begin{equation}
\sqrt{\frac{3(2N\!-\!\bar{n})}{4N}}\!\!\left(\!\!e^{i\theta_1}\!\!\ket{\frac{N}{3},N\!\!}
\!+\!e^{i\theta_2}\!\!\ket{N,\frac{N}{3}\!}\!\!\right)\!+\!\sqrt{\frac{3\bar{n}\!-\!4N}{2N}}\!\!\ket{N\!N}.
\end{equation}

These OFPSs are analytically obtained in the ideal case, namely, without considering the existence of any noise. Their performance 
under the noise of particle loss has been thoroughly studied and compared to other finite-dimensional states like the N00N state 
and twin-Fock state in Ref.~\cite{Qin2025}. When the noise of particle loss exists, the aforementioned OFPSs may no longer be optimal 
anymore in mathematics. In the optimization process, the diagonalization of $\rho_{\phi}$ is usually required, which makes it difficult to 
search the OFPS analytically in this case. Hence, numerical methods are the first choice in this work to locate the true OFPS under 
particle loss. 

\section{True OFPS under particle loss}

%======================================= Figure ======================================
\begin{figure}[tp]
\centering
\includegraphics[width=7.5cm]{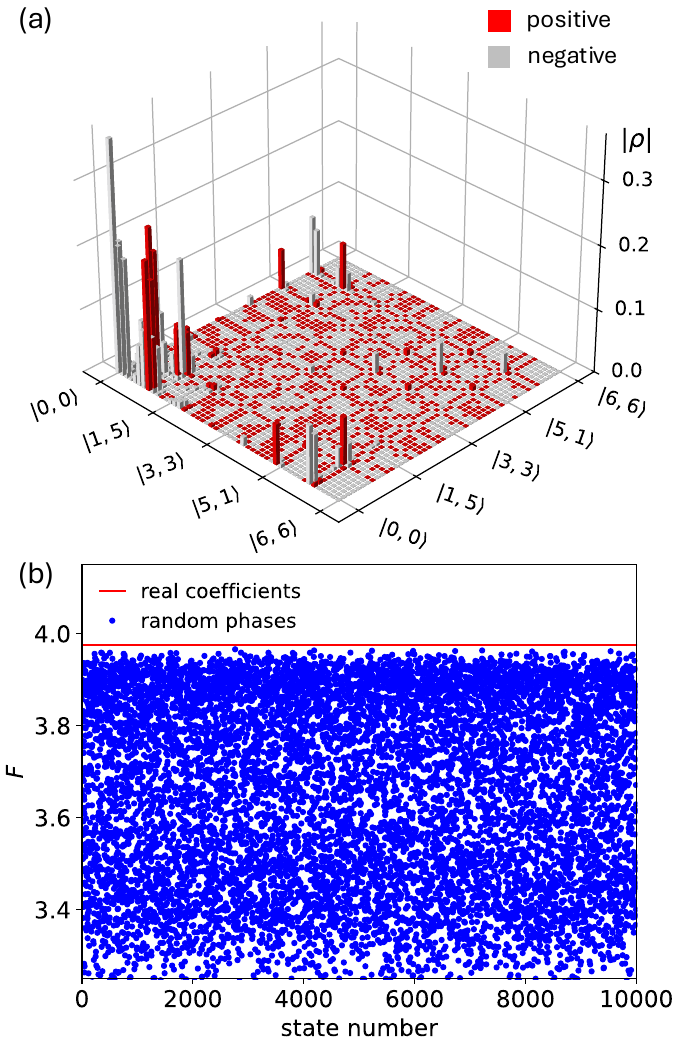}
\caption{(a) Tomography of the noisy OFPS obtained via COBYLA for $T_1=T_2=0.8$ 
in the case of $\bar{n}=2$ and $N=6$. The red and gray bars represent that the value 
of the corresponding entry is positive and negative, respectively. (b) Validity test 
of the noisy OFPS obtained with real coefficients. 10 thousand groups of random phases 
(blue dots) are generated and used for the test. The red line represents the QFI for 
the noisy OFPS with real coefficients.}
\label{fig:tomo}
\end{figure}
%===================================================================================

In finite-dimensional probe states optimization, the constraint of a fixed input average particle number $\bar{n}$ 
makes it a constrained optimization problem. For the finite-dimensional state given in Eq.~(\ref{eq:general_form}), 
the constrained optimization problem can be formulated as
\begin{eqnarray}
& \underset{C_{ij}}{\mathrm{max}} & ~F(C_{ij}) \nonumber \\
& \text{s.t.} & \begin{cases}
|C_{ij}|\in [0,1], \forall i,j\in [0,N], \\
\sum^N_{i,j=0} |C_{ij}|^2=1,  \\
\sum^N_{i,j=0} |C_{ij}|^2(i+j)=\bar{n},  
\end{cases} 
\end{eqnarray}
where the symbol "s.t." is short for "subject to". Hence, in this case an efficient constrained
optimization algorithm is required. Fortunately, Powell~\cite{Powell1994,Powell1998,Powell2007} developed a very
powerful algorithm, the constrained optimization by linear approximation (COBYLA) algorithm, which fits the problem
in our case. For the sake of simplification of the problem, here we take $C_{ij}$ as a real coefficient to reduce 
the number of the variables in the optimization. 

%======================================= Figure ======================================
\begin{figure*}[tp]
\centering
\includegraphics[width=18.0cm]{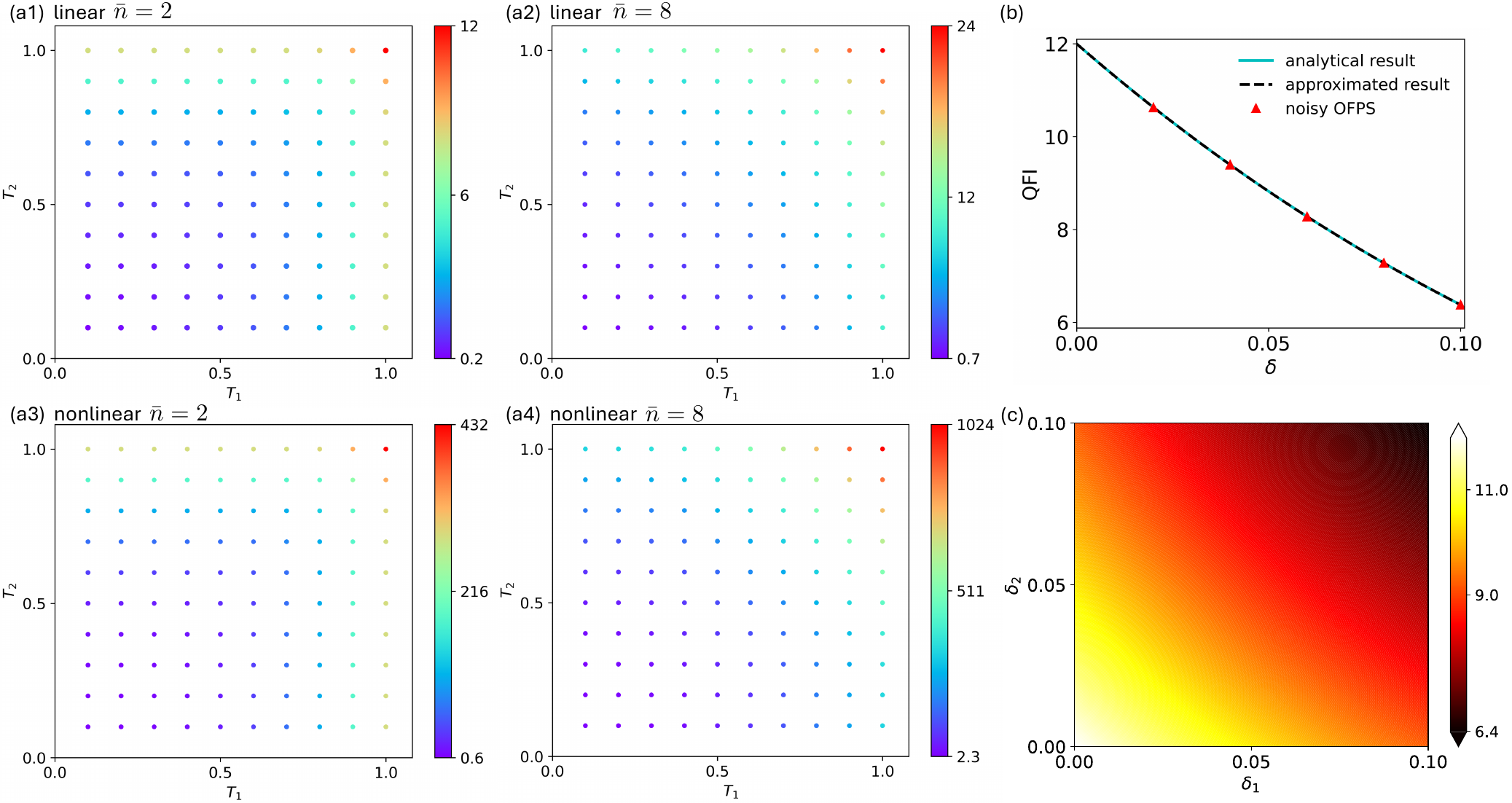}
\caption{The values of the QFI for the true OFPSs under particle loss as a function of 
transmission coefficients $T_1$ and $T_2$ for (a1) linear phase shifts with $\bar{n}=2$ 
($\bar{n}<N$); (a2) linear phase shifts with $\bar{n}=8$ ($\bar{n}>N$); (a3) nonlinear 
phase shifts with $\bar{n}=2$ ($\bar{n}<N$); and (a4) nonlinear phase shifts with 
$\bar{n}=8$ ($\bar{n}>N$). (b) The values of QFI for the analytical (solid cyan line) and 
approximated (dashed black line) results regarding to the noiseless OPFS and that 
of the noisy OFPSs (red triangles) for different values of $\delta$ ($\delta_1=\delta_2=\delta$). 
(c) The behaviors of the approximated QFI with respect to the noiseless OFPS given in 
Eq.~(\ref{eq:lin_Optstate}) as a function of $\delta_1$ and $\delta_2$ in the linear case 
with $\bar{n}=2$. $N=6$ and $\phi_{\mathrm{true}}$ is set to be 0.2 in all plots.}
\label{fig:T}
\end{figure*}
%===================================================================================

The COBYLA algorithm is a gradient-free optimization method for constrained nonlinear optimization problems~\cite{Powell1994,
Powell1998,Powell2007}. The core philosophy of this algorithm is the linear approximations of the objective and constraint 
functions at the vertices of the simplices via linear interpolation, and then solving the approximated linear optimization 
problem with simplex methods. This algorithm can be directly invoked with the function \emph{minimize(method='COBYLA')} or 
\emph{fmin\_cobyla()} in the Python package SciPy. Moreover, this state optimization scenario will be integrated into the 
package QuanEstimation~\cite{Zhang2022,Yu2025} soon. 

Utilizing the COBYLA algorithm, the true OFPS under particle loss can be identified. As a demonstration, the located noisy 
OFPS for $T_1=T_2=0.8$ in the case of $\bar{n}=2$ and $N=6$ is illustrated in Fig.~\ref{fig:tomo}(a), which shows that the 
noiseless OFPS indeed ceases to be optimal anymore. The numerical optimization finishes when the variety of the optimized 
QFI converges to the scaling of $10^{-6}$. Furthermore, random phases are added to the optimized coefficients to test the 
validity of the real optimal coefficients. As shown in Fig.~\ref{fig:tomo}(b), 10 thousand groups of random phases are 
generated and added to the real optimal coefficients. The corresponding QFIs for these states (blue dots) are no larger 
than the one with real optimal coefficients (red line). This fact indicates that the optimal solution indeed exists for real 
coefficients in this case. In fact, in a specific case one can combine these two steps to locate the noisy OFPS with complex 
coefficients. 

It is not difficult to realize that the formula of the noisy OFPS relies on the values of the transmission coefficients. To 
investigate the behavior of the maximum precision limit with different transmission coefficients, $100$ noisy OFPSs are 
obtained via the COBYLA algorithm in both linear and nonlinear cases with $N=6$ and $\bar{n}=2,8$. The program 
was run on a cluster with each node 56 CPUs (2.7\,GHz) and 240\,G memory. The values of the QFI with respect to the noisy OFPSs in 
the regions of $\bar{n}<N$ ($\bar{n}=2$) and $\bar{n}>N$ ($\bar{n}=8$) are given in Figs.~\ref{fig:T}(a1) and \ref{fig:T}(a2) for 
the linear phase shifts, and Figs.~\ref{fig:T}(a3) and \ref{fig:T}(a4) for the nonlinear phase shifts. The separation of the regions 
is due to the existence of different formulas of the noiseless OFPS in these two regions. In all figures the negative effect of 
particle loss coincides with the general understanding that it becomes more significant when the values of $T_1$ and $T_2$ reduce. 
Since the formulas of the noisy OFPSs are usually more complex than their noiseless counterparts, the preparation of the noisy OFPSs 
might also be more complex in practice. Hence, a more important question here is when the noiseless OFPSs can still present comparable 
performance with the noisy OFPSs. 

It is obvious that the comparable performance can only exist when the particle loss is not significant. Hence, the QFI for large transmission 
coefficients need to be investigated for the sake of answering the aforementioned question. Denote $T_1=1-\delta_1$ and $T_2=1-\delta_2$ 
with $\delta_1$ and $\delta_2$ two small quantities, then the Taylor expansion of the $i$th eigenvalue and eigenstate of $\rho$, up to the 
first order of $\delta_1$ and $\delta_2$, can be expressed by 
\begin{align}
\lambda_0 &=1-\delta_1\lambda_{0,1}-\delta_2\lambda_{0,2}, \\    
\lambda_i &=\delta_1\lambda_{i,1}+\delta_2\lambda_{i,2}, \forall i\neq 0, 
\end{align}
and 
\begin{align}
\ket{\lambda_0} &=\ket{\psi_{\mathrm{in}}}+\delta_1\ket{\lambda_{0,1}}+\delta_2\ket{\lambda_{0,2}}, \\    
\ket{\lambda_i} &=\delta_1\ket{\lambda_{i,1}}+\delta_2\ket{\lambda_{i,2}}, \forall i\neq 0. 
\end{align}
Here $\lambda_{0,1}$, $\lambda_{i,1}$ ($\lambda_{0,2}$, $\lambda_{i,2}$) and $\ket{\lambda_{0,1}}$, $\ket{\lambda_{i,1}}$ ($\ket{\lambda_{0,2}}$, 
$\ket{\lambda_{i,2}}$) are the first order terms with respect to $\delta_1$ ($\delta_2$). The existence of $\lambda_0$ and $\ket{\lambda_0}$ is 
due to the fact that $\rho$ is actually the pure state $\ket{\psi_{\mathrm{in}}}$ when $T_1=T_2=1$. Based on Eq.~(\ref{eq:QFI}), the QFI with 
respect to $\rho$ up to the first order (denoted by $F_{\mathrm{lin}}$) in the linear case is 
\begin{equation}
F_{\mathrm{lin}}=4\lambda_0\left(\bra{\lambda_0}J^2_z\ket{\lambda_0}-\bra{\lambda_0}J_z\ket{\lambda_0}^2\right), 
\end{equation}
namely, the approximated QFI is just the QFI with respect to $\ket{\lambda_0}$ multiplying $\lambda_0$. 
When the probe state $\ket{\psi_{\mathrm{in}}}$ is the noiseless OFPS given in Eq.~(\ref{eq:lin_Optstate}) or (\ref{eq:lin_Optstate1}), the term 
$\bra{\psi_{\mathrm{in}}}J_z\ket{\psi_{\mathrm{in}}}$ vanishes, which means 
\begin{equation}
F_{\mathrm{lin}}\approx 4\lambda_0\left(\bra{\lambda_0}J^2_z\ket{\lambda_0}\right).
\end{equation}

This fact can be further confirmed by Fig.~\ref{fig:T}(b), where the behaviors of the analytical (solid cyan line) and approximated (dashed 
black line) results coincide with each other in the case of symmetric loss ($\delta_1=\delta_2=\delta$) with $\bar{n}=2$ and $N=6$. More 
importantly, the noiseless OFPS presents comparable performance with the noisy OFPSs (red triangles), hence, in this region of loss the 
noiseless OFPS would be a better choice as the probe state. Moreover, this result also indicates that in this region 
both the optimality and performance of the noiseless OFPS are very robust. As to the robustness of the noisy OFPSs in the entire region, 
it is difficult to define the robustness in this case since the form of the noisy OFPS relies on the loss rate. Different loss rates usually 
correspond to different noisy OFPSs. Therefore, further investigations are still needed to quantify the robustness against the loss in general.

\begin{figure}[tp]
\centering
\includegraphics[width=8.cm]{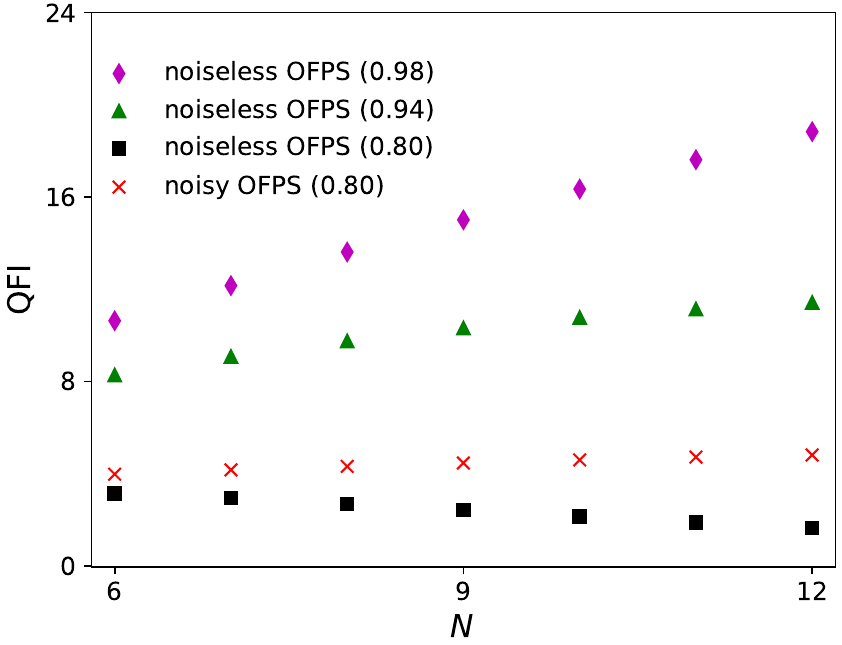}
\caption{The QFIs of the noiseless and noisy OFPSs with different values of the Fock dimension $N$. 
The purple diamonds, green triangles, black squares represent the QFIs of the noiseless OFPSs for the transmission 
coefficients $T_1=T_2=0.98, 0.94, 0.80$, respectively. The red crosses represent the QFIs of the noisy OFPSs with 
respect to $T_1=T_2=0.8$. In the entire figure $\bar{n}=2$. }
\label{fig:OFPSvsN}
\end{figure}

Apart from the average particle number, the Fock dimension is another powerful resource in the OFPS to further improve the precision 
limit in the absence of noise~\cite{Qin2025}. However, the impact of loss on the effectiveness of the Fock dimension is still unknown. 
Due to the previous discussions, the noiseless OFPSs present comparable performance in the case that $T_1=T_2\geq 0.9$, hence, in 
this region the noiseless OFPSs can be used instead of the noisy OFPSs to reveal the influence of the particle loss on the effectiveness 
of the Fock dimension. As shown in Fig.~\ref{fig:OFPSvsN}, the QFIs of the noiseless OFPSs in the cases of $T_1=T_2=0.98$ (purple 
diamonds) and $0.94$ (green triangles) increase as the Fock dimension grows, indicating that Fock dimension is still an effective 
resource to improve the precision limit in this region. In the case that $T_1=T_2=0.8$, the QFIs of the noiseless OFPSs (black squares) 
decrease as the Fock dimension grows. This fact means that if the noiseless OFPSs are still taken as the probe states in this case, Fock 
dimension ceases to be an effective resource. If the noisy OFPSs are applied, the QFIs (red crosses) recover to a slowly growth trend.  
Hence, in this case the Fock dimension is still an effective resource when the noisy OFPSs are used. However, its efficiency on the 
performance improvement is limited.

%================================================ Figure ===============================================
%=====================================================================================================

\section{Optimal measurement}

%================================================ Figure ===============================================
\begin{figure*}[tp]
\centering
\includegraphics[width=16cm]{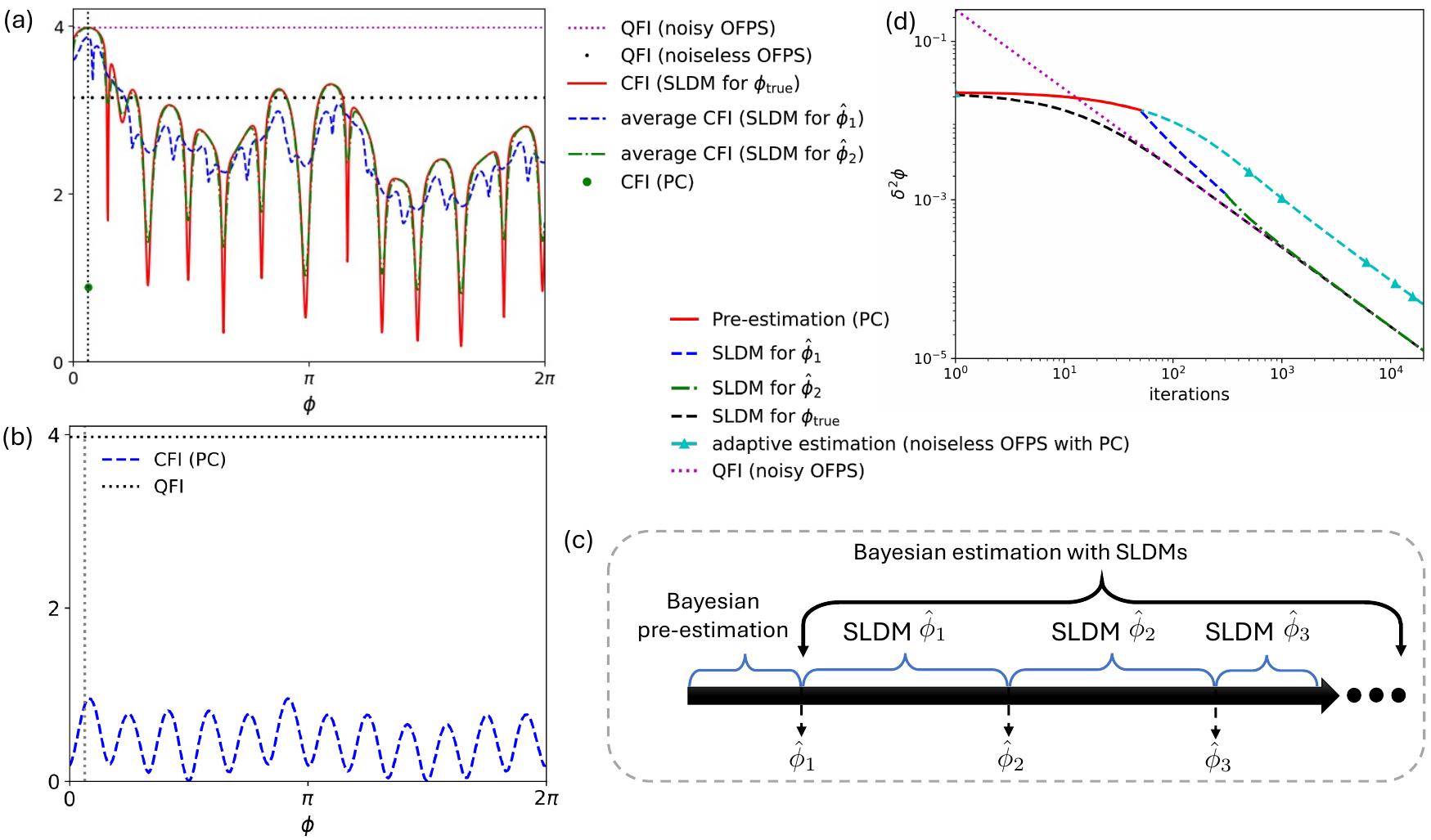}
\caption{(a) The behaviors of the QFI and CFI (or average CFI) for noisy and noiseless OFPSs 
as functions of phase difference $\phi$. The average CFIs of the noisy OFPS correspond to the 
SLDM with different estimated values, and are the average of 2000 simulations; (b) The 
values of the CFI and QFI for the noisy OFPS with PC measurement in the entire region of $\phi$; (c) 
Illustration of the two-step strategy to realize the optimal measurement without the knowledge of the 
true value; (d) Simulation of the performance of the two-step strategy in practice. All lines are the average 
performance of 2000 simulations. In (a), (b) and (d) the parameters are set as $\bar{n}=2$, $N=6$, and 
$T_1=T_2=0.8$. The true value $\phi_{\mathrm{true}}=0.2$. }
\label{fig:OptM}
\end{figure*}
%=====================================================================================================

A complete protocol for quantum phase estimation not only contains the probe state, but 
also the measurement. To fit the usual form of the measurement in quantum interferometry, 
a beam splitter is added before the measurement is executed, as illustrated in Fig.~\ref{fig:schematic}. 
The theoretical operator for this beam splitter is $\exp(i\pi J_x/2)$ with $J_x=(a^{\dag}b+ab^{\dag})/2$ 
a Schwinger operator. 

With the absence of noise, both parity and particle-counting measurements are optimal to the OFPS for some specific 
true values~\cite{Qin2025}. The dependency of the optimality of these two measurements on the true value $\phi_{\mathrm{true}}$ 
can be overcome by the adaptive measurement, where a tunable phase is involved and tuned properly. When the particle loss exists, 
both the parity and particle-counting measurements cease to be optimal with respect to the noisy OFPS, and even the noiseless 
OFPS. As a demonstration, the QFIs of the noisy OFPS (purple dots) and noiseless OFPS (black dots), and the CFI (green circle) of 
the particle-counting (PC) measurement with respect to the noisy OFPS are illustrated in Fig.~\ref{fig:OptM}(a) in the case 
of $\bar{n}=2$ and $N=6$. The transmission coefficients are $T_1=T_2=0.8$ and the true value $\phi_{\mathrm{true}}$ is 
set to be 0.2. It can be seen that in this case the noiseless OFPS indeed ceases to be optimal due to the large gap between 
its QFI and that of the noisy OFPS. In the meantime, the PC measurement is not an optimal measurement to the noisy OFPS. 
The CFI of the parity measurement is no larger than that of the PC measurement, and thus is not optimal either. 
Furthermore, although the performance of the adaptive PC measurement can reach the maximum CFI in the entire 
region of $\phi$,  it is still far from the value of the QFI,  as shown in Fig.~\ref{fig:OptM}(b), which means the performance of the 
adaptive measurement in this case is limited. Therefore, the optimal measurement of the noisy OFPS has to be located for the sake 
of providing a complete optimal scheme. 

One simple way to construct the optimal measurement is using the eigenvectors of the SLD. Denote $\ket{L_k}$ as the $k$th 
eigenvector of $L$, then the projective measurement (SLDM) at the point of the true value (denoted by 
$\{e^{i\frac{\pi}{2}J_x}\ketbra{L_k}e^{-i\frac{\pi}{2}J_x}\}_{\phi=\phi_{\mathrm{true}}}$, the rotation of $e^{i\frac{\pi}{2}J_x}$ 
is used to cancel the effect of the second beam splitter) is the optimal measurement, i.e., the CFI of the SLDM at this point equals 
the QFI. Here the CFI can be calculated via the equation 
\begin{equation}
I=\sum_k\frac{\left(\partial_{\phi}\bra{L_k}\rho_{\phi}\ket{L_k}\right)^2}{\bra{L_k}\rho_{\phi}\ket{L_k}}.     
\end{equation}
The CFI of the SLDM in the case of $\bar{n}=2$ and $N=6$ is also illustrated in Fig.~\ref{fig:OptM}(a) (red line) as a demonstration. 
At the point of $\phi_{\mathrm{true}}$, as shown in the plot, the CFI of the SLDM reaches the QFI, indicating that it is indeed optimal 
at this value point. 

Similar to the parity and PC measurements in the noiseless scenario, SLDM depends on the phase difference $\phi$ and is only optimal 
at the point $\phi=\phi_{\mathrm{true}}$. However, in practice $\phi_{\mathrm{true}}$ is not known, and more importantly, the choice of 
$\phi$ in the SLDM does not only affect the optimality like the parity and PC measurements, but also affects the specific form of the 
SLDM. Hence, when the particle loss exists we have to find a way to realize the optimal measurement without the knowledge of 
$\phi_{\mathrm{true}}$. 

In this paper, we provide a two-step measurement strategy to realize the optimal measurement in practice without a high prior knowledge 
of the true value before the experiment, as illustrated in Fig.~\ref{fig:OptM}(c). The first step is the Bayesian pre-estimation. Both the parity 
and PC measurements can be used in this step. The PC measurement is usually more efficient and requires fewer iteration numbers. After 
the Bayesian pre-estimation, the estimated value $\hat{\phi}_1$ is obtained and used to construct the SLDM, i.e., 
$\{e^{i\frac{\pi}{2}J_x}\ketbra{L_k}e^{-i\frac{\pi}{2}J_x}\}_{\phi=\hat{\phi}_{1}}$. The second step is the Bayesian estimation with the SLDM. 
In this step, several SLDMs are constructed with the estimated values of $\phi$. For example, after proper 
iterations of the Bayesian estimation with $\{e^{i\frac{\pi}{2}J_x}\ketbra{L_k}e^{-i\frac{\pi}{2}J_x}\}_{\phi=\hat{\phi}_{1}}$, a new estimated value 
$\hat{\phi}_{2}$ is obtained and used to construct the second SLDM $\{e^{i\frac{\pi}{2}J_x}\ketbra{L_k}e^{-i\frac{\pi}{2}J_x}\}_{\phi=\hat{\phi}_{2}}$, 
which is further applied in the estimation to obtain another new estimated value. This process repeats until the precision converges. The iteration 
numbers of the pre-estimation and each SLDM need to be determined case by case in practice. Due to the limited performance of the Bayesian 
estimation with the parity or PC measurement under particle loss, the pre-estimation does not need too many iterations. The improvement would 
be very little once it converges. 

The case of $\bar{n}=2$ and $N=6$ is still taken to demonstrate the performance of the given two-step strategy. The simulation of the experiment 
is given in Fig.~\ref{fig:OptM}(d). The PC measurement is chosen in the first step. The prior probability distribution of $\phi$ is a uniform one in the 
region $[0,\pi/6]$. After 50 iterations of pre-estimation (solid red line), $\hat{\phi}_{1}$ is obtained. In the second step, two SLDMs are constructed 
and applied. The first one is constructed via $\hat{\phi}_1$ and used in 250 iterations (dashed blue line); The second one is constructed via 
$\hat{\phi}_{2}$, which is obtained at the end of the 300th iteration, and used in the rest iterations (dash-dotted green line). All lines are the average 
of 2000 simulations. It shows that the result of the two-step strategy coincides with the theoretical optimal measurement 
$\{e^{i\frac{\pi}{2}J_x}\ketbra{L_k}e^{-i\frac{\pi}{2}J_x}\}_{\phi=\phi_{\mathrm{true}}}$ (dashed black line, also the average of 2000 simulations).  
And it reaches the precision limit given by the QFI (dotted purple line). This performance is way better than the adaptive PC measurement of the 
noiseless OFPS (solid-triangle cyan line, the average of 2000 simulations), and more technical details of the Bayesian and adaptive measurements 
can be found in Ref.~\cite{Qin2025}. To more intuitively understand the two-step strategy, the average CFIs of the SLDMs with respect to 
$\hat{\phi}_1$ and $\hat{\phi}_2$ are given in Fig.~\ref{fig:OptM}(a) as the dashed blue and dash-dotted green lines. It can be found that the 
values of the average CFI increase at the point of $\phi_{\mathrm{true}}$ when the measurement changes from 
$\{e^{i\frac{\pi}{2}J_x}\ketbra{L_k}e^{-i\frac{\pi}{2}J_x}\}_{\phi=\hat{\phi}_{1}}$ to 
$\{e^{i\frac{\pi}{2}J_x}\ketbra{L_k}e^{-i\frac{\pi}{2}J_x}\}_{\phi=\hat{\phi}_{2}}$. More importantly, the performance of the SLDM 
$\{e^{i\frac{\pi}{2}J_x}\ketbra{L_k}e^{-i\frac{\pi}{2}J_x}\}_{\phi=\hat{\phi}_{2}}$ basically coincides with the theoretical optimal measurement 
$\{e^{i\frac{\pi}{2}J_x}\ketbra{L_k}e^{-i\frac{\pi}{2}J_x}\}_{\phi=\phi_{\mathrm{true}}}$, especially at the point of $\phi_{\mathrm{true}}$. 
Hence, the given two-step strategy can indeed realize the optimal measurement of the noisy OFPS in practice and reach the ultimate precision limit. 

\section{Conclusion}

In conclusion, in this paper we have located the OFPSs for quantum phase estimation under the noise of particle loss. The performance of the 
noisy OFPSs are investigated for different amounts of particle loss. The noiseless OFPSs show a comparable performance with the noisy OFPSs 
in the case of small losses. 

Optimal measurement is indispensable for a complete optimal phase estimation scheme. Different from the noiseless case, the parity and 
particle-counting measurements cease to be optimal under the noise of particle loss. The projective measurement constructed via the 
eigenvectors of the symmetric logarithmic derivative (SLDM) is then applied as the optimal measurement instead. To solve the problem 
that the dependence of the SLDM on the true value, which makes it unrealizable in practice, a two-step strategy based on the Bayesian 
estimation is proposed. Utilizing the numerical simulation of the practical experiment, this strategy is proved to be an efficient way to realize 
the ultimate precision limit quantified by the quantum Fisher information. 

Quantum interferometry with the OFPSs provides a full new approach, i.e., Fock dimension, to further improve the phase sensitivity without 
increasing the particle number of the probe. To make the OFPSs applicable in practice, their preparations in various physical platforms are 
crucial and require deep investigations. Given the current quantum information techniques, quantum circuits appear to be the most promising 
platform for the practical implementation of OFPSs in the near future and worth further investigation. Moreover, the forms of the OFPSs for 
quantum multiphase estimation are still open questions and also worth to be further studied. 

Particle loss is a common noise mode in quantum interferometry, yet not the only one. The preparation inefficiency of the 
probe state and the detector inefficiency are also very important noise sources~\cite{Wiseman1997,Datta2011,Frascella2021}. The effect of 
the preparation inefficiency is similar to the particle loss, which also makes the state mixed before it goes into the phase shifts. Hence, the 
OFPSs under this noise can also be readily solved via the COBYLA algorithm. The involvement of detector inefficiency would be more completed. 
The form of the OFPS is not affected by the detector inefficiency in principle since the used objective function is the QFI. However, it is highly 
possible that the detector inefficiency would make it hard or even impossible to find the optimal measurement to attain the QFI of the OFPS. 
An alternative approach is to optimize the probe state and measurement simultaneously with the CFI as the objective function. The performance 
of this optimization process is hard to predict due to the general difficulty of multivariable optimizations. Therefore, the optimal phase estimation 
schemes with finite-dimensional states involving the detector inefficiency is still an open problem and needs to be further investigations in the 
future. In the meantime, locating the noisy OFPSs poses a significant computational challenge for the COBYLA algorithm when the Fock dimension 
is large. State-of-the-art learning algorithms might be potential candidates to solve this problem.

Due to the importance of quantum interferometry in quantum information and technology, the proposed phase estimation scheme with the OFPSs 
would be very useful in practice. It has great potential to be applied soon in both quantum industry and fundamental research like gravitational-wave 
detections and tests of fundamental physics. 

\begin{acknowledgments}
This work was supported by the National Natural Science Foundation of China (Grant Nos.\,12575013, 12547103 and 12175075). 
\end{acknowledgments}

\end{document}